\documentclass{ws-procs975x65}
\usepackage{graphicx}
\def\beq{\begin{equation}}
\def\eeq{\end{equation}}
\newcommand{\aj}{{AJ}}%
\newcommand{\apj}{{Ap.~J.}}%
\newcommand{\apjl}{{Astrophys. J. Lett.}}%
\newcommand{\apjs}{{Astrophys. J. Suppl. Ser.}}%
\newcommand{\aap}{{A\&A}}%
\newcommand{\mnras}{{MNRAS}}%
\newcommand{\prd}{{Phys. Rev.~D}}%
\newcommand{\pasp}{{Publ. Astron. Soc. Pac.}}%
\newcommand{\ssr}{{Space Sci. Rev.}}%
\newcommand{\nat}{{Nature}}%
\begin{document}

\title{Observational  properties of magnetic white dwarfs}
\author{Lilia Ferrario$^*$}

\address{Mathematical Sciences Institute, The Australian National University,\\
Canberra, ACT 2601, Australia\\
$^*$E-mail: Lilia.Ferrario@anu.edu.au\\
www.maths.anu.edu.au/people/lilia-ferrario}

\begin{abstract}

  There are no known examples of magnetic white dwarfs with fields larger than
  $\sim 3$\,MG paired with a non-degenerate companion in detached
  binary systems.  The suggestion is that highly magnetic, isolated
  white dwarfs may originate from stars that coalesce during common
  envelope evolution while those stars that emerge from a common
  envelope on a close orbit may evolve into double degenerate systems
  consisting of two white dwarfs, one or both magnetic.

  The presence of planets or planetary debris around white dwarfs is
  also a new and exciting area of research that may give us important
  clues on the formation of first and second generation planetary
  systems, since these place unique signatures in the spectra of white
  dwarfs.

\end{abstract}


\bodymatter


\section{Introduction}

Grw$+70^\circ8247$ \cite{Kuiper1934} was the first magnetic white
dwarf (MWD) discovered. This object exhibits a strongly circularly
polarised spectrum that is nearly featureless except for broad shallow
features that became known as ``Minkowski bands''.  Computations of H
transitions in strong magnetic fields in the mid-80s revealed that
these features are Zeeman shifted H lines in a field of $100-320$\,MG
\cite{Wickramasinghe1988}. From just a handful of objects that were
known in the early 80s, the number has increased to $\sim 300$
isolated MWDs and $\sim 170$ MWDs in binaries \cite{FMG2015}.

The modelling of Zeeman spectral lines is the best way to determine
the field strength and structure of isolated MWDs. The detection of
cyclotron lines in the optical to IR bands offers an additional method
for the study of MWD fields in interacting binaries
\cite{Ferrario92,Ferrario1993,Ferrarioetal93,Schmidt2001}.  Whilst
research on these enigmatic objects has led to a better understanding
of stellar magnetism, it has also raised a number of unanswered
questions regarding the pre-MWD evolution and the origin and structure
of stellar magnetic fields.

In this brief review I will report the latest thoughts on the origin
and evolution of MWDs and I will highlight their importance as
possible progenitors of supernova events, millisecond pulsars,
and as hosts of first and second generation planets.  More
comprehensive reviews on all aspects of this subject are available in
the literature \cite{FMG2015,FMZ2015}.

\section{Origin of magnetic fields in white dwarfs}

\begin{figure}
\centering
  \includegraphics[width=\textwidth]{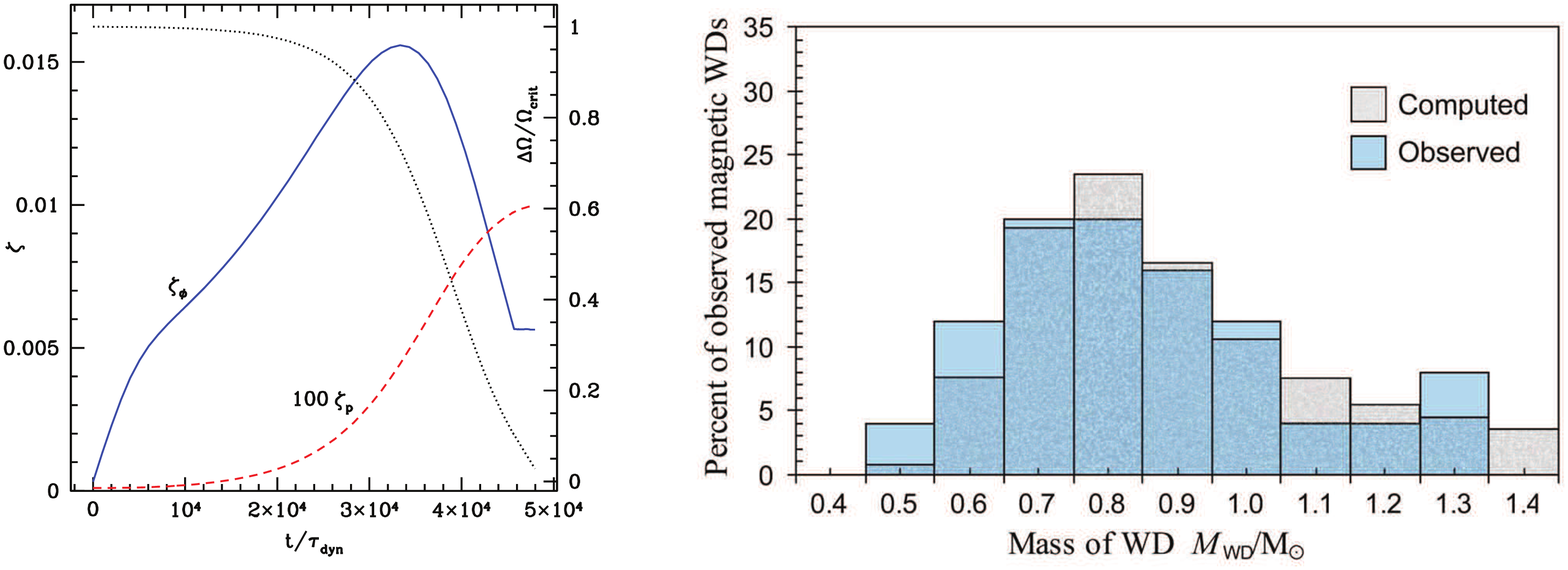}
\caption{Left panel: Evolution of $B_{\rm p}$ (poloidal field) and $B_\phi$
  (toroidal field, $\zeta = \sqrt{\eta}\approx B$) during a star
  merging event is shown on the left hand axis. Here,
  $B_\phi$ reaches a peak and then decays till equilibrium with the
  poloidal field is reached. The decay of differential rotation
  $\Delta\Omega$ is shown on the right hand axis \cite{Wick14}. Right
  panel: Mass distribution of observed MWDs (blue histogram) compared
  with the computed sample (grey histogram) \cite{Briggs2015}. The most common
  predicted progenitors are Asymptotic Giant Branch (AGB)/MS star
  (61\%), AGB/M-dwarf (17\%) and RGB/WD (14\%) mergers. Other
  routes comprise less than 1 per cent of the total.}
\label{Bs_Mass}
\end{figure}

According to the fossil field scenario, magnetic fields present in the
interstellar medium (ISM) at the time of star formation freeze into
the radiative regions of proto-stars. Fluctuations in the ISM field
strength would explain the field spread observed in main sequence (MS)
stars.  In this picture the magnetic flux is then conserved as the
star evolves off the MS to the compact star phase.  Since the peculiar
Ap/Bp stars are known to have large-scale ordered fields, these stars
have been proposed to be the progenitors of the MWDs \cite{Wick05}.
However, one should expect that MWDs ought to occur as often in
detached binaries as in single stars.  The Sloan Digital Sky Survey
has identified thousands of WD+M dwarfs in detached spectroscopic
binaries \cite{Rebassa13}.  The peculiarity is that none of these
systems harbours a MWD.  The sample of WDs within 20 pc has revealed
that 19.6$\pm$4.5\% of them have MS companions \cite{Holberg2009}. The
magnitude limited Palomar Green (PG) Survey has shown
that $23-29$\% of hot WDs have cool companions \cite{Holberg2005}.  Thus, the current MWD
sample of about 300 stars should contain at least $40-90$ objects with
a non-degenerate companion \cite{Liebert05,Liebert2015}, but instead
it does not include any.  This incongruity suggests that the origin of
magnetic fields in WDs may be linked to their duplicity
\cite{Tout2008}. Spiralling stellar cores during common envelope (CE)
evolution give rise to differential rotation. Differential rotation
and a seed poloidal magnetic field in the envelope of the primary star
would produce strong toroidal and poloidal fields which would
stabilise each other and limit field growth.  The final poloidal field
strength has been found to be proportional to the initial amount of
differential rotation, but independent of the initial seed field
\cite{Wick14} (see left panel of Fig. \ref{Bs_Mass}).

MWDs in non-interacting double degenerate systems (DDs) and in the
magnetic cataclysmic variables could form via a similar mechanism
during CE evolution.  The closer the stellar cores get before the
expulsion of the envelope, the stronger the magnetic field of the
proto-MWD emerging from the CE will be\cite{Tout2008}.

It has been known since 1988 \cite{Liebert1988} that high field MWDs have masses that are
on average higher than those of their non-magnetic or weakly magnetic counterparts
($\sim 0.78$\,$M\odot$ \cite{FMG2015} versus $\sim 0.66$\,$M\odot$
\cite{Tremblay2013}). Binary population synthesis calculations are
consistent with the scenario that MWDs arise from stars that merge
during CE evolution \cite{Briggs2015} and predict that MWDs formed
through this channel are on average more massive than non-magnetic
white dwarfs, as required by observations (see right panel of Fig. \ref{Bs_Mass}).

\begin{figure}[h]
\begin{center}
\includegraphics[width=\textwidth]{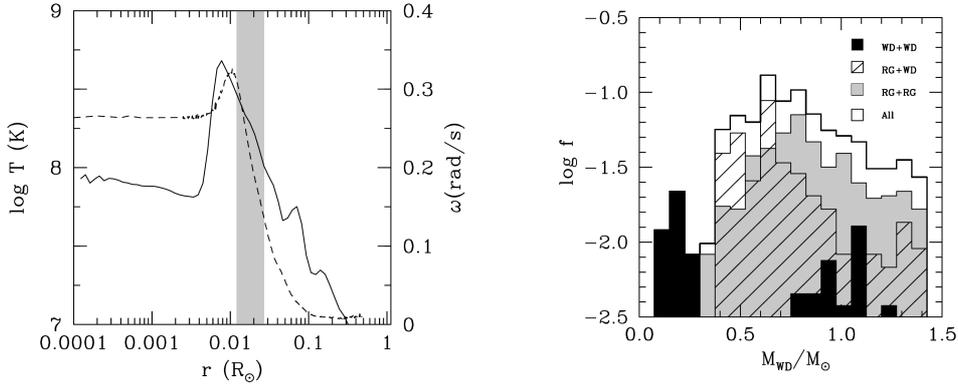}
\end{center}
\caption{Left panel: Dynamo configuration in the merger of two stars
  of 0.6\,M$_\odot$ and 0.8\,M$_\odot$. The temperature profile (solid
  line) is shown on the left hand axis and the rotational velocity (dashed line)
  on the right axis as a function of the radius of the resulting WD. Right
  panel: Mass frequency of merger channels \cite{garcia2012}. Black
  histogram: white dwarf-white dwarf mergers; dashed histogram: red
  giant-white dwarf mergers; shaded histogram: red giant-red giant
  mergers.  The solid line represents the total mass distribution.}
\label{Garcia-Berro2012}
\end{figure}

In an alternative scenario, seed fields in the accretion disc formed
by the debris of a low-mass star tidally disrupted by its companion
during CE, are augmented by turbulence and shear and transferred on to
what will become a MWD \cite{Nordhaus2011}. 3-D hydrodynamic
calculations of merging double WDs have also been carried out
\cite{garcia2012}. These studies show that a differentially rotating,
hot, convective corona forming around the more massive star during the
merging process produces strong magnetic fields (see left panel of
Fig. \ref{Garcia-Berro2012}). Population synthesis computations of
double WD mergers appear to be consistent with observations (see right
panel of Fig. \ref{Garcia-Berro2012}).

On the other hand one should not completely dismiss the fossil field
scenario for the formation of MWDs \cite{Tout2004,Wick05}, since it is
feasible that MWDs could exist in binaries although hidden in the
glare of a brighter non-degenerate companion \cite{Ferrario2012}
(Sirius-type systems).  Thus, the lack of MWD+M dwarfs in detached
binaries could perhaps be attributed to the manner in which magnetic
Ap/Bp stars (the progenitors of the MWDs in the fossil field
hypothesis) are paired with their secondary stars.  If a sufficient
number of Sirius-type systems hosting a MWD is discovered, this
finding could support the view that there is more than one formation
channel for MWDs. In this context, I wish to draw attention to the
atmospheric modelling of cool MWDs \cite{Rolland2015} showing that the
spectral characteristics of 10 out of the 16 objects under
consideration in this study can be best modelled as unresolved
magnetic DDs (see section \ref{DoubleDeg}) whose progenitors could
indeed have been (magnetic) Sirius-type systems.

\section{Double degenerate systems}\label{DoubleDeg}

Type Ia supernovae (SNeIa), which are used as standard candles to
determine the expansion history of the universe, are caused by
accreting white dwarfs which undergo thermonuclear explosions.  Since
different SNIa progenitors may lead to different peak luminosities or
different light-curve width-luminosity relations, their
characteristics as a function of redshift could imitate the behaviour
of a time-varying dark energy equation of state
\cite{Linder2006}. Since we still know very little about the origin of
SNeIa, this is known as the ``SNIa progenitor puzzle''.

The thermonuclear disruption of a WD could occur through the merger of
two WDs (also called the DD route).  On the other hand, a DD merger
may also lead to the formation of a millisecond pulsar via accretion
induced collapse \cite{Hurley2010,Ferrario2007}. If one of the two WDs
has a strong magnetic field, a collapse could lead to the formation of
a high field pulsar or a magnetar \cite{Ferrario2006,Ferrario2008}.
What causes a collapse rather than an explosion is still being
strongly debated. The presence of an intense magnetic field has only
very recently been included in simulations of the post-merger
evolution of DDs \cite{Ji2013}. A list of the currently known magnetic
DDs is given in Table \ref{tab_DD} below.

LB\,11146 was identified in 1993 \cite{Liebert1993} as a DD binary
consisting of a non-magnetic and a strongly magnetic ($\sim 670$\,MG)
WD. The WD masses were estimated to be about $0.9 M_\odot$ and thus,
LB\,11146 was considered to be a potential SNIa progenitor. However,
further observations of LB\,11146 \cite{Nelan2007} revealed the
presence of a long orbital period ($\sim 130$\,days) and a separation
of $\sim 0.6$\,AU.  Since the size of the orbit is smaller than the
size of its AGB progenitor, this indicated that the two stars went
through at least one CE phase. The long orbital period of LB\,11146 is
at odds with those of most other known DDs \cite{Nelan2007}, whose
periods vary from a few hours to a few days \cite{Nelemans2005}.

Another remarkable DD is NLTT\,12758, whose MWD component has a polar
field of $\sim 3$\,MG. This DD exhibits variations at the rotation
(spin) period of the MWD ($\sim 22$\,mins) and at the orbital period ($\sim
11$\,hrs) \cite{Kawka2012,Kawka2016}. The best fit model gives masses
of $\sim 0.9$ and $\sim 1$\,M$_\odot$ for the non-magnetic and
magnetic components respectively. As for LB\,11146, the total mass
exceeds the Chandrasekhar's limiting mass, so in principle this DD
could be a potential SNIa progenitor. However, its orbital parameters
also place this DD on a merging timescale comparable to the Hubble
time \cite{Kawka2016}.

Follow-up observations of EUVE\,J1439+750 \cite{Vennes2015},
G\,183-35\cite{Brinkworth2013}, and G\,141-2 \cite{Brinkworth2013}
have shown that there is no evidence for variability in any of these
objects, at least within the given observational limitations. More
observations are needed to establish the orbital parameters of the
remaining magnetic DDs.

\begin{table}
\tbl{Currently known magnetic DDs }
{\begin{tabular}{@{}llll@{}}
\toprule
Name & Other name &Variability/Periods   &  Notes \& References \\
\colrule
0040+000 &SDSS\,J004248.19+001955.3 &$\cdots$ &\refcite{Schmidt2003}\\
0121-429  & LHS\,1243                           &$\cdots$ &\refcite{Rolland2015}\\
0239+109 &G\,4-34, LTT\, 0886              &$\cdots$ &\refcite{Gianninas2011}\\
0325-857 &RE\,J0317-853, EUVE\,J0317-855 & $P_{\rm spin}\sim725$\,s, $P_{\rm orb}\sim 2092$\,yrs & \refcite{Vennes2003,Farihi2008,Lawrie2013}\\
0410-114 &G160-51, NLTT\,12758&$P_{\rm spin}\sim 23$\,min, $P_{\rm orb}\sim 0.6$\,d & \refcite{Kawka2012,Rolland2015,Kawka2016}\\
               &SDSS\,J092646.88+132134.5  &$\cdots$ &CPM, \refcite{Dobbie2012}\\
0512+284& LSPM\,J0515+2839               &$\cdots$ &\refcite{Rolland2015}\\
0745+303& SDSS\,J074853.07+302543.5 & $\cdots$ &CPM, \refcite{Dobbie2013}\\
                & SDSS\,J092646.88+132134.5& $\cdots$& CPM, \refcite{Dobbie2012}\\
0945+246& LB\,11146                            & $P_{\rm orb}\sim 130$\,d& \refcite{Nelan2007}\\
1026+117& LHS\,2273                            &$P_{\rm spin?}\sim35-45$\,min& \refcite{Brinkworth2013,Rolland2015}\\
1300+590& SDSS\,J130033.48+590407.0 &$\cdots$ &CPM, \refcite{Girven2010}\\
1330+015& G\,62-46                              &$\cdots$& \refcite{Rolland2015}\\
1440+753 &EUVE\,J1439+750                  &$\cdots$ & \refcite{VennesFerrWick1999}\\
1503-070 &GD\,175                                &$\cdots$& \refcite{Rolland2015}\\
               &SDSS\, J150746.80+520958.0  &$\cdots$&CPM, \refcite{Dobbie2012}\\
1506+399 &CBS\,229, SDSS\,J150813+394504 &$\cdots$& CPM, \refcite{Dobbie2013}\\
1514+282& SDSS\,J1516+2803               &$\cdots$& \refcite{Rolland2015}\\
1713+393& NLTT\,44447                       &$\cdots$& \refcite{Rolland2015}\\
1814+248& G\,183-35                          &50\,min--yrs?&Double MWDs, \refcite{Putney1997,Brinkworth2013,Rolland2015}\\
1818+126 &G\,141-2                            &yrs? & \refcite{Brinkworth2013,Nelan2015,Rolland2015}\\

\botrule
\end{tabular}
}
\label{tab_DD}
\end{table}

According to the merger scenario for magnetic field formation
\cite{Tout2008}, in non-interacting WD-MWD binaries that have gone
through CE evolution, the non-magnetic WD would form first following a
CE phase that would bring the two stars closer to each other. The
absence of a magnetic field implies that there must be additional and
as yet unknown conditions that need to be satisfied for a magnetic
field to be generated during a CE (e.g. stellar separation, mass
ratio). It is during a subsequent CE phase that the companion star
would develop a magnetic field and would evolve into a MWD. Those DDs
with large binary separations (e.g. common proper motion pairs) that
did not undergo CE evolution would be the results of triple systems
where the MWD formed via a CE merger. There is at least one DD,
G\,183-35, where both WDs are magnetic. This system could have formed
following two CE phases each producing a MWD. If this is the case, the
implication is that the components of G\,183-35 form a close binary
system and that further observations may be needed to search for very
short orbital periods.

Finally, one needs to address the question why not all DDs,
particularly those at small binary separations, harbour at least one
MWD. The answer could be that either (i) the WD precursor entering CE
evolution did not already have a degenerate core \cite{Briggs2015}; or
(ii) convective nuclear burning in the core of the WD precursor took
place after it emerged from CE evolution, thus destroying any
generated frozen-in magnetic field \cite{Briggs2015} or, as already
mentioned above, (iii) there are additional factors that come into
play for the generation of a magnetic field when two stars enter a CE
phase.

According to the alternative fossil field hypothesis, magnetic DDs
would be the descendants of Sirius-type systems \cite{Ferrario2012}
where the MWD in the binary would the progeny of a magnetic Ap/Bp star.

\section{Planets around MWDs}\label{planets}

\begin{figure}[h]
\begin{center}
\includegraphics[width=0.7\textwidth]{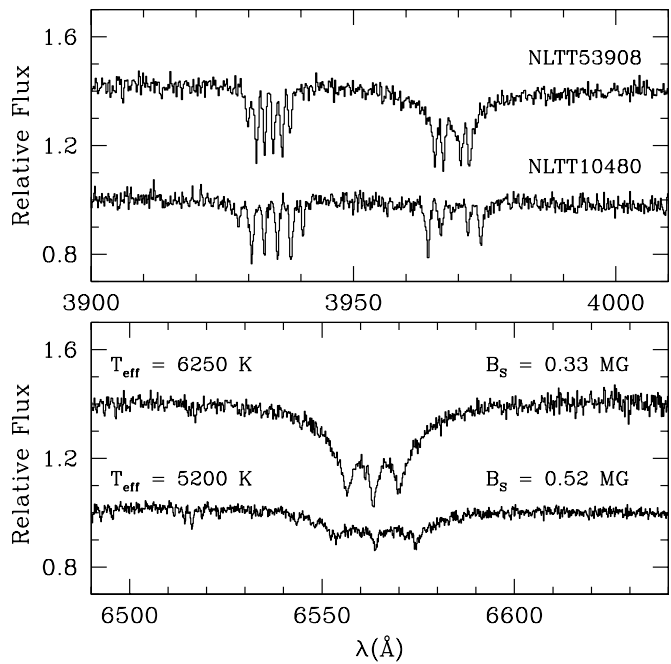}
\end{center}
\caption{Linear Zeeman splitting of Ca II (H and K) and
  H$\alpha$ in NLTT\,10480\cite{Kawka2011} and NLTT\,53908
  \cite{Kawka2014}. }
\label{DZ_MWD}
\end{figure}

The accretion of planetary debris by WDs could explain the existence
of DZ WDs\cite{Zuckerman2003} which are characterised by strong metal
lines (e.g. Ca, Si, Mg, Na and sometimes Fe, Ti and Cr) in their
spectra (see Fig. \ref{DZ_MWD}).
 Interestingly, the incidence of magnetism among
cool ($< 9,000$\,K) DZ WDs is about 13\%, which is considerably higher
than among the general population of WDs \cite{Kawka2014,Hollands2015}
(magnitude-limited samples indicate that only $\sim 2-5$\% are
magnetic \cite{Kepler2015}). In order to explain this peculiarity, it
has been suggested that the progenitor AGB star underwent a CE phase
with a giant gaseous planet, which led to the formation of a MWD
\cite{Farihi2011}. The atmosphere of this MWD would then become
polluted due to the accretion of rocky planets and/or asteroid
rubble\cite{Farihi2010,Zuckerman2010}. The CE formation channel for DZ MWDs could be
strengthened if the incidence of magnetism among DC (featureless) WDs,
which belong to the same stellar population as the DZ's
\cite{Farihi2010}, were found to be considerably different.  In this
picture, the MS progenitors of DZ MWDs could have been intermediate
mass stars that had planetary systems composed by Earth-like planets
and at least one close-in gaseous giant responsible for the WD's
magnetic field \cite{Farihi2010}. 

It is still rather unclear what the incidence of magnetism among DZ
WDs with $T_{\rm eff}\ge10,000$\,K is \cite{Kawka2014}, although
preliminary studies seem to indicate that there is a dearth of warm DZ
MWDs. This may indicate that magnetic fields in WDs with first or
second generation planetary systems could be created a few Gigayears
after the formation of the WD \cite{Farihi2011}. The problem with this
scenario is that the incidence of large gaseous planets with short
enough orbital periods to cause a merger and thus field production in
the WD, may be far too low to explain the high percentage of magnetism
among DZ WDs.

Another unique object that cannot be forgotten when discussing planets
around WDs is GD\,356 with a field strength of $\sim13$\,MG. GD\,356
exhibits Zeeman triplets of H$\alpha$ and H$\beta$ in emission. The
modelling of the spectral features of this star has shown that the
emission lines are formed in a region covering a tenth of the stellar
surface where the stellar atmosphere has an inverted temperature
distribution \cite{Ferrario1997_GD356}. Radio observations have failed
to provide arguments in support of a magnetic active corona and IR
observations to search for a low-mass stellar companion have also
given no results \cite{Wickramasinghe2010}. In order to explain these
observations, it has been argued that this MWD may have a planetary
companion in a close orbit that was stripped-down to its iron core
\cite{Li1998} during the post-MS evolution. Electrical currents
flowing between the MWD and the planet would cause the heating, via
ohmic dissipation, of the upper layers of the MWD's atmosphere. The
loss of energy due to Ohmic dissipation would then cause the orbit to
decay \cite{Li1998}. The model's prediction was that GD\,356 should
exhibit periodic variations of several hours. Such a variability was
indeed detected in 2004 \cite{Brinkworth2004} and found to be
consistent with the modulation of a spot located near the spin axis
and covering $\sim10$\% of the stellar surface. This electric current
model is also known as ``the unipolar inductor model''
\cite{Willes2005}.
\begin{figure}[h]
\begin{center}
\includegraphics[width=\textwidth]{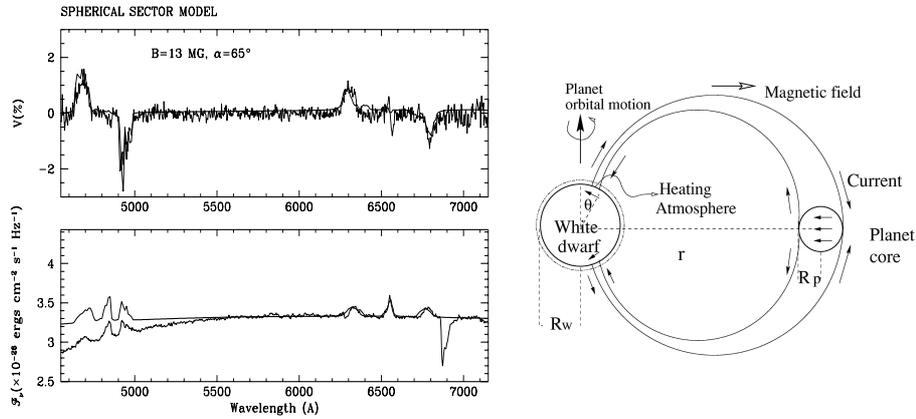}
\end{center}
\caption{Left panel: Observed intensity (lower panel) and polarisation
  (upper panel) spectra of GD\,356 showing its characteristic emission
  features. The calculated spectra are superimposed to the data. Right
  panel: Electric currents are generated by the highly conductive iron
  core of a planet around the MWD. The electric circuit that is setup
  heats the MWD's atmosphere in a region near its magnetic pole causing 
  temperature inversion and the formation of emission lines. }
\label{GD356}
\end{figure}

Thus, WDs can be effectively used to gain insight on how planets form
around Sun-like stars, which are their progenitors. In this
context, I draw attention on new research on Earth-like planets around
cool WDs \cite{Agol2011,Fossati2012}. The authors theorise about how
planets could stay in the habitable zone of a cool WD for as long as
8\,Gyr. They also discuss how planets could end up in such a zone and
propose that they could form from material orbiting the WD, possibly
as a result of binary interaction or merger events
\cite{Livio2005,Wickramasinghe2010}. This process would be similar to that invoked to
explain the presence of planets originating from accretion discs
formed after supernova events around neutron stars.
\cite{Wolszczan1992,Phinney1993}. These planets are often called
``second generation planets''. Alternatively, planets may end up
orbiting the WD via capture or through migration from outer to inner
orbits \cite{Debes2002}. This kind of formation scenario would be
applicable to any planet found in a close orbit around a WD star.

\section*{Acknowledgments}

The author wishes to thank her collaborators, Dayal Wickramasinghe,
Chris Tout, Stephane Vennes and Adela Kawka for many long and
stimulating discussions.

\end{document}